\documentclass{elsart}
\usepackage{epsfig}
\begin{document}
\begin{frontmatter}


\title{Exotic Quantum Order in Low-Dimensional Systems}
\author{S.M. Girvin}
\address{Department of Physics,
Swain Hall West 117,
Indiana University,
Bloomington, Indiana 47405 USA}


\begin{abstract}
Strongly correlated quantum systems in low dimensions often exhibit
novel quantum ordering.  This ordering is sometimes hidden and can be
revealed only by examining new `dual' types of correlations.  Such
ordering leads to novel collective modes and fractional quantum numbers.
Examples will be presented from quantum spin chains and the quantum Hall
effect.
\end{abstract}
\end{frontmatter}

\section{Introduction}

The last two decades have witnessed remarkable experimental discoveries 
of novel
quantum phenomena in strongly correlated systems in low dimensions.  Along with
this has come a vast increase in our understanding of how to describe the
 underlying order in these systems which often remains hidden to ordinary
probes
 and has required the invention of new types of order
 parameters.  As examples of this I will discuss here the Haldane gap in
quantum
 spin chains, the Laughlin gap in the fractional
quantum Hall effect and  novel types of ordering in quantum Hall ferromagnets.

The underlying theme will be that a state which appears disordered in one
representation can appear ordered in a dual representation.
  The word duality will be used in the sense of  ordinary `wave--particle'
duality
 in quantum mechanics, and in the technical field-theoretic
 sense referring to a transformation which interchanges particles and vortices,
or
interchanges charge and flux.

\section{Quantum Spin Chains}

Let us begin with the idea of `wave--particle' duality for quantum spins.
Quantum spins are strange objects which behave in some ways like
 ordinary classical angular momentum vectors which can point in arbitrary
directions.
  Yet if we measure their projection along
 any given direction, we find that it can take on only a discrete set of $2S+1$
 values.  The discreteness suggests that a collection of quantum
 spins might exhibit discrete excitations and a gap.  The continuous
orientation
 picture suggests a continuous spectrum of gapless spin waves. To be 
specific, let us
 consider the case of spin 1/2.  The two possible discrete states 
can
be represented by the spinors $|\uparrow\rangle$, and
$|\downarrow\rangle$.  However, quantum mechanics allows for
the possibility of a coherent linear superposition of these two states
\begin{equation}
|\psi\rangle = \cos(\theta/2) e^{i\varphi/2} |\uparrow\rangle +
\sin(\theta/2) e^{-i\varphi/2} |\downarrow\rangle.
\label{eq:coherent}
\end{equation}
This continuous family of states is parameterized by the two Euler angles
determining the orientation of the expectation value of the spin vector.

Spin 1/2 is the least classical possible spin value, and yet the continuous
spin wave picture works extremely well in dimensions greater than one.  
In the antiferromagnet the staggered magnetization
\begin{equation}
\vec M_s \equiv \sum_j (-1)^j \langle \vec S_j\rangle
\end{equation}
is non-zero in the ground state (for $d>1$).  
There exists a continuous family of 
spin wave
excited states in which the Euler angles precess at a frequency linearly
proportional to the wave vector (at long wavelengths).  The same holds 
true for
all possible other spin values $S = 1,3/2,2,\ldots$  They are all effectively
equivalent and the discreteness of the individual spin orientations is
irrelevant to the low energy physics.

\begin{figure}[thb]
\centerline{\epsfxsize 4in\epsfbox{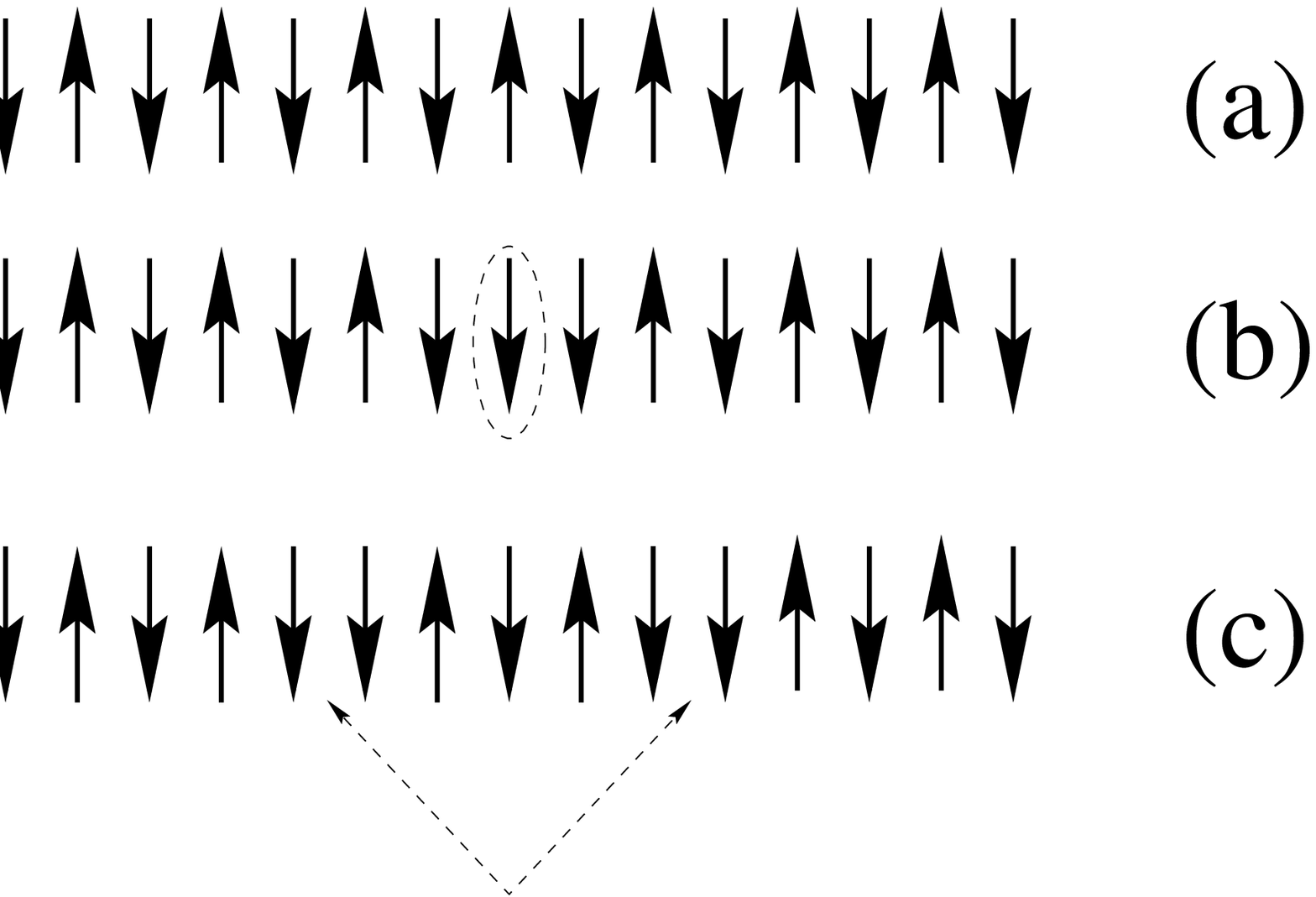}}
\caption{
(a) staggered ordered state in 1D. (b) excitation caused by flipping a
single spin in the center of the chain. (c) fractionalization of the
single flipped spin into two $S=1/2$ domain walls (denoted by dashed
arrows).  The system moves from state (b) to state (c) by mutual spin
flip of the pairs of spins on each side of the central spin.
}
\label{fig:spinon1}
\end{figure}

The situation is very different in one dimensional spin chains where quantum
fluctuations destroy the long range staggered magnetic order even at zero
temperature.  Half-integer spins $S=1/2,3/2,5/2,\ldots$
still have gapless excitations, but they are  {\em not} spin waves.    It turns
out that the spin wave, which would normally represent a $\Delta S=1$ 
excitation
above the singlet ground state, splits up (`fractionalizes') into two
independent fermion-like objects known as `spinons'.  These objects are
actually domain walls in the staggered order as shown in
Fig.(\ref{fig:spinon1}).  This process is readily detectable experimentally
because an ordinary spin wave is undamped  at long wavelengths and so has a
single sharply defined energy associated with each wavelength.  A pair of
spinons on the other hand has much more phase space available to it and
the
spectral density is a convolution of the spectral densities of the individual
spinons.  Neutron scattering data demonstrating the existence of spinons is
presented in Ref.~\cite{spinon-expt}.

\begin{figure}[thb]
\centerline{\epsfxsize 4in\epsfbox{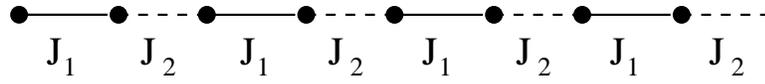}}
\caption{
Dimerized $S=1/2$ chain with alternating couplings $J_1 > J_2$.  The solid
lines indicate singlet bonds.
}
\label{fig:dimer}
\end{figure}

\begin{figure}[thb]
\centerline{\epsfxsize 4in\epsfbox{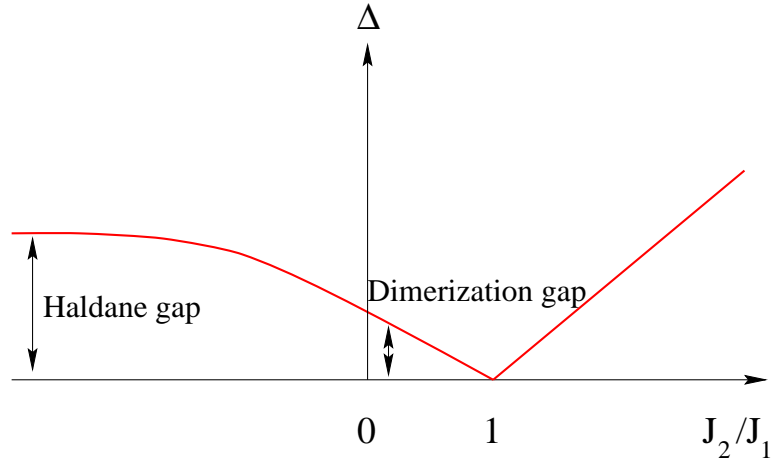}}
\caption{
Excitation gap $\Delta$   for a dimerized
$S=1/2$ chain as a function of the bond strength ratio $J_2/J_1$.  The
Haldane $S=1$ chain corresponds to $J_2 \longrightarrow -\infty$.
}
\label{fig:dimer_gap}
\end{figure}

For integer spin values $S=1,2,3,\ldots$,
the discreteness becomes relevant and the spin chain exhibits the so-called
Haldane excitation gap. This can be understood by means of the following
construction which relates the $S=1$ chain  to a dimerized  $S=1/2$ chain.
Fig.(\ref{fig:dimer})  shows a spin chain in which the exchange coupling 
alternates
between two different values $J_1$
and $J_2$ with  $J_1 > J_2$.  In order to capture the essence of the ground
state let us consider the special case $J_2=0$.  Here we know the exact ground
state consists of  independent singlets on the solid bonds illustrated in
Fig.(\ref{fig:dimer}).  This is a `valence bond solid'.  It has  
a non-degenerate
ground state and a highly degenerate first excited state in which (any)
 one of the
singlets is converted into a triplet.  It costs a finite energy to break the
singlet bond and so there is an excitation gap.  If we now turn the second
coupling $J_2$ back on, the valence bonds begin to fluctuate but the 
essential
character of the state remains unchanged.  The singlets are more likely to be
found on the $J_1$ links, and the excitation gap is stable against the $J_2$
perturbation.  The gap remains open for any value of $J_2<J_1$  as illustrated in Fig.(\ref{fig:dimer_gap}).

\begin{figure}[thb]
\centerline{\epsfxsize 4in\epsfbox{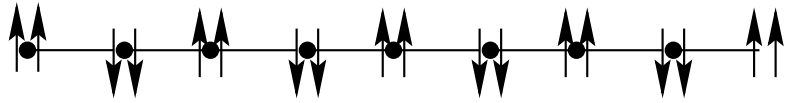}}
\caption{
Valence bond solid state in which the $S=1$ spins on each site are
represented as a symmetrical combination of two $S=1/2$ objects.  Each
of these objects is paired into a singlet with one of its neighbors.
}
\label{fig:vbs}
\end{figure}

Fig.(\ref{fig:dimer_gap})  also shows that the dimerization gap remains 
open even as the
value of $J_2$ passes through zero and becomes negative.  In the limit
$J_2\longrightarrow -\infty$,
the pairs of spins on the $J_2$ links are forced to become parallel and their
resulting triplet becomes an effective spin one degree of freedom, as shown 
in
Fig.(\ref{fig:vbs}). The dimerization gap has now become the Haldane gap 
for an integer
spin chain.  The resulting `valence bond solid' is an apparently featureless
gapped spin `liquid' state.  Because of the excitation gap $\Delta$, the local
spin correlation function decays exponentially in (imaginary) time
\begin{equation}
\langle S^z_j(\tau) S^z_j(0)\rangle \sim e^{-\tau/\Delta}.
\end{equation}
Similarly, the equal-time spin correlation function decays exponentially in
space
\begin{equation}
\langle S^z_j S^z_{j+r}\rangle \sim e^{-r/\xi}.
\end{equation}

This  apparent featurelessness  obscures an underlying hidden order   which is
characteristic of the valence bond solid state.  It turns out that if one
examines a typical configuration of the spins in the ground state it looks 
like this:
\begin{equation}
+00-+0000-+-0+00-+-000+0-
\end{equation}
  Notice that if we ignore the spins with $S^z=0$,
the remaining spins have perfect \cite{caveat} antiferromagnet order,
\begin{equation}
+-+-+-+-+-+-+-+-
\end{equation}
Because of the random number of $S^z=0$  sites inserted between the $S^z=\pm 1$
 sites, the long-range antiferromagnetic order is invisible to the ordinary
spin-spin correlation function which, as noted above, decays exponentially.
This `hidden' order is however manifest in the non-local den Nijs `string'
correlation function \cite{dennijs,arovas-smg,kohmoto} defined by
\begin{equation}
\mathcal{O}^z_s \equiv \langle S^z_j \exp{\left\{i\pi\sum_{k=j+1}^{j+r-1}
S^z_k\right\}}S^z_{j+r}\rangle.
\end{equation}
This object exhibits long-range order in the Haldane gapped phase
\begin{equation}
\lim_{r\longrightarrow\infty} \mathcal{O}^z_s \ne 0.
\end{equation}
Breaking this topological order to create an excitation costs a finite energy
gap.

\begin{figure}[thb]
\centerline{\epsfxsize 4in\epsfbox{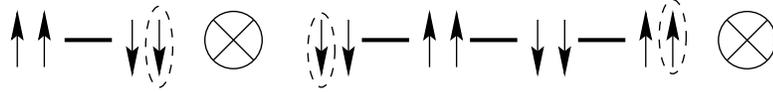}}
\caption{
Valence bond solid interrupted by non-magnetic impurities.  The missing
valence bonds liberate a \textit{fractional} spin ($S=1/2$) degree of
freedom (indicated by the dashed ellipses) 
at the ends of each broken chain segment.
}
\label{fig:impurities}
\end{figure}

Because of the excitation gap, the spin degrees of freedom disappear at low
temperature and the magnetic susceptibility becomes exponentially small.  One
of the remarkable and paradoxical features of the Haldane phase is that
introduction of {\em nonmagnetic}  impurities actually liberates spin degrees
of freedom.  These are not, as one might have naively expected, $\Delta S = 1$
excitations but rather pairs of {\em  fractional spin} $\Delta S = 1/2$
excitations.  The process is illustrated in Fig.(\ref{fig:impurities}).  
Recall that the spin 1 on each site is viewed as a pair of spin 1/2 
objects bound ferromagnetically into a triplet.  As noted earlier, the 
valence bond solid ground state (in the absence of disorder) has each of 
the $S=1/2$ objects pairing into a singlet bond with one of the spin 1/2 
objects on the neighboring sites. (This  is done in a symmetric way so
that 
each site has total spin precisely equal to unity.)  If one of the 
neighbors is a non-magnetic impurity, one of the singlet bonds is broken 
and {\em one-half} of the spin on the site is liberated.  
These weakly interacting fractional spins remain free to low temperatures 
and produce a Curie-like power law tail in the magnetic susceptibility.  
This is an example of a quantum McCoy-Griffiths singularity in which the 
den Nijs  order (may) still be present but the gap is 
not \cite{hymanthesis,hyman-yang}.

\section{Fractional Quantum Hall Effect}

The fractional quantum Hall 
effect \cite{dassarmabook,smgbook,chakraborty,stone}
(FQHE) is easily one of the richest and most remarkable phenomena 
discovered in many years.  It turns out that there are deep analogies to 
the $S=1$ spin chains we have just examined.  Instead of an apparently 
featureless gapped {\em spin} liquid, we have an apparently featureless 
gapped {\em charge} liquid.
As we shall see there is also a close analog of the hidden non-local string
order.  Breaking this order with a topological defect liberates a fractional
{\em charge} (and in some cases spin as well).

For simplicity, let us focus on the $\nu=1/3$ Hall plateau characterized by
conductivity $\sigma_{xx} = 0, \sigma_{xy} = \frac{1}{3}\frac{e^2}{h}$.  The
Hamiltonian for the two-dimensional electron gas is
\begin{equation}
H = \sum_j \frac{1}{2m}\left(\vec p_j + \frac{e}{c} \vec A_j\right)^2
+ \frac{1}{2} \sum_{i<j} v(\vec r_i - \vec r_j),
\end{equation}
where $\vec A_j \equiv \vec A(\vec r_j)$ is the vector potential for the strong
external magnetic field.  To discover the hidden order in the Laughlin 
ground state we
must make a singular gauge transformation which attaches three
 flux quanta to each
electron.  The Aharonov-Bohm phases associated with these flux tubes 
introduces an extra minus sign upon particle exchange and converts these 
composite objects
into bosons \cite{chap10,ODLRO,Read,Kivelson-Zhang}. The physics is 
completely
identical in this representation, but the Hamiltonian changes to
\begin{equation}
H_{\rm CB} = \sum_j \frac{1}{2m}\left(\vec p_j + \frac{e}{c} (\vec A_j+\vec
a_j) \right)^2
+ \frac{1}{2} \sum_{i<j} v(\vec r_i - \vec r_j),
\end{equation}
where
\begin{equation}
\vec\nabla\times\vec a_j = - \sum_{\ell\ne j} 3\Phi_0 \delta^2(\vec r_j - 
\vec r_l)
\label{eq:11}
\end{equation}
is the attached flux and $\Phi_0$ is the flux quantum.  The sign of the
attached flux has been chosen so that on the average, the attached flux cancels
the external field
\begin{equation}
\langle \vec\nabla\times(\vec a_j+\vec A_j)\rangle = 0.
\end{equation}
Thus in the mean field approximation we have composite bosons moving in zero
magnetic field.  It is the condensation of these bosons that is the essential
ordering in the quantum Hall effect.  Because these objects behave as if they
carry 2D Coulomb charge, they exhibit algebraic (rather than true)
off-diagonal long-range order
\begin{equation}
\langle \psi_{\rm CB}(\vec r) \psi^\dagger_{\rm CB}(\vec r')\rangle \sim
|\vec r - \vec r'|^{-3/2}.
\end{equation}
If we undo the singular gauge transformation to express this correlation 
function
in terms of the original fermions, we have
\begin{equation}
\langle \psi_{\rm F}(\vec r)e^{i\int_{\vec r}^{\vec r'}
d\vec R\cdot \vec a(\vec R)} \psi^\dagger_{\rm F}(\vec r\,'\,)\rangle .
\end{equation}
Because, as can be seen from eq.(\ref{eq:11}),
 $\vec a$ depends on the position of {\em all} the particles, this
non-local object is closely analogous to the `string' correlator that 
describes
the hidden order in integer spin chains.

Let us now examine the nature of topological defects (vortices) in the
composite boson condensate.  It turns out that because the objects obey an
effective 2D electrodynamics (with logarithmic  interactions among the
charges), we expect flux quantization as in a superconductor
\begin{equation}
\int d^2r\, \vec\nabla\times(\vec a+\vec A) = \pm \Phi_0.
\end{equation}
However  because of the flux attachment transformation we have
\begin{equation}
\vec\nabla\times(\vec a+\vec A) = 3\Phi_0 \delta\rho
\end{equation}
where $\delta\rho$ is the density deviation away from the mean.  Hence, flux
quantization implies that vortices carry fractional charge \cite{chap10}
\begin{equation}
\int d^2r\, \delta\rho = \pm \frac{1}{3},
\end{equation}
and thus are the Laughlin quasiparticles.  These are analogous to the
fractional spin excitations  appearing at the ends of $S=1$ chains.
(A more precise analogy can be made with the gapless excitations at the
edges of quantum Hall liquids, which are in a sense a gas of Laughlin
quasiparticles liberated at the edge \cite{wen,kane-fisher-book}.)

Another regime occurs at filling factor $\nu=1/2$.  This can the analyzed by
attaching two rather than three flux quanta to the electrons, converting them
not into composite bosons but rather composite {\em fermions} moving in zero
average magnetic 
field. \cite{HLR,jain-book-review,halperin-book-review,fradkin} 
While there are strong fluctuation corrections to 
this mean field picture which are not yet fully understood, the basic 
phenomenology suggested by the picture has received striking experimental
confirmation. \cite{willett,kang}

\section{Quantum Hall Ferromagnets}

We turn now to the subject of ferromagnetism in quantum Hall systems, an area
where significant advances have been recently made \cite{smg-ahmbook}.
 The large external
magnetic field couples very strongly to the orbital motion and quenches the
kinetic energy into discrete Landau levels.  It turns out that the external
field couples rather weakly to the spin degrees of freedom and so low energy
spin fluctuations are not completely
frozen out by the (small) Zeeman splitting.

We will consider here the case of a single filled Landau level.  The ground
state is fully spin polarized because this makes the spatial wave function
antisymmetric which minimizes the Coulomb energy.  We can describe the
spin configuration in low-lying excited states buying introducing a unit vector
field $\vec m(\vec r)$ to describe the local spin orientation.  The energy 
for slowly varying spin textures must then be of the form
\begin{equation}
U=\frac{1}{2}\rho_s\int d^2r\, \partial_\mu
 m^\nu \partial_\mu m^\nu - h \int d^2r\,  m^z.
\end{equation}
The spin stiffness $\rho_s \sim 5$K is a measure of how strongly the exchange
energy prefers for the spins to be locally parallel and $h\sim 2$K  
represents the Zeeman coupling.

\begin{figure}[thb]
\centerline{\epsfxsize 4in\epsfbox{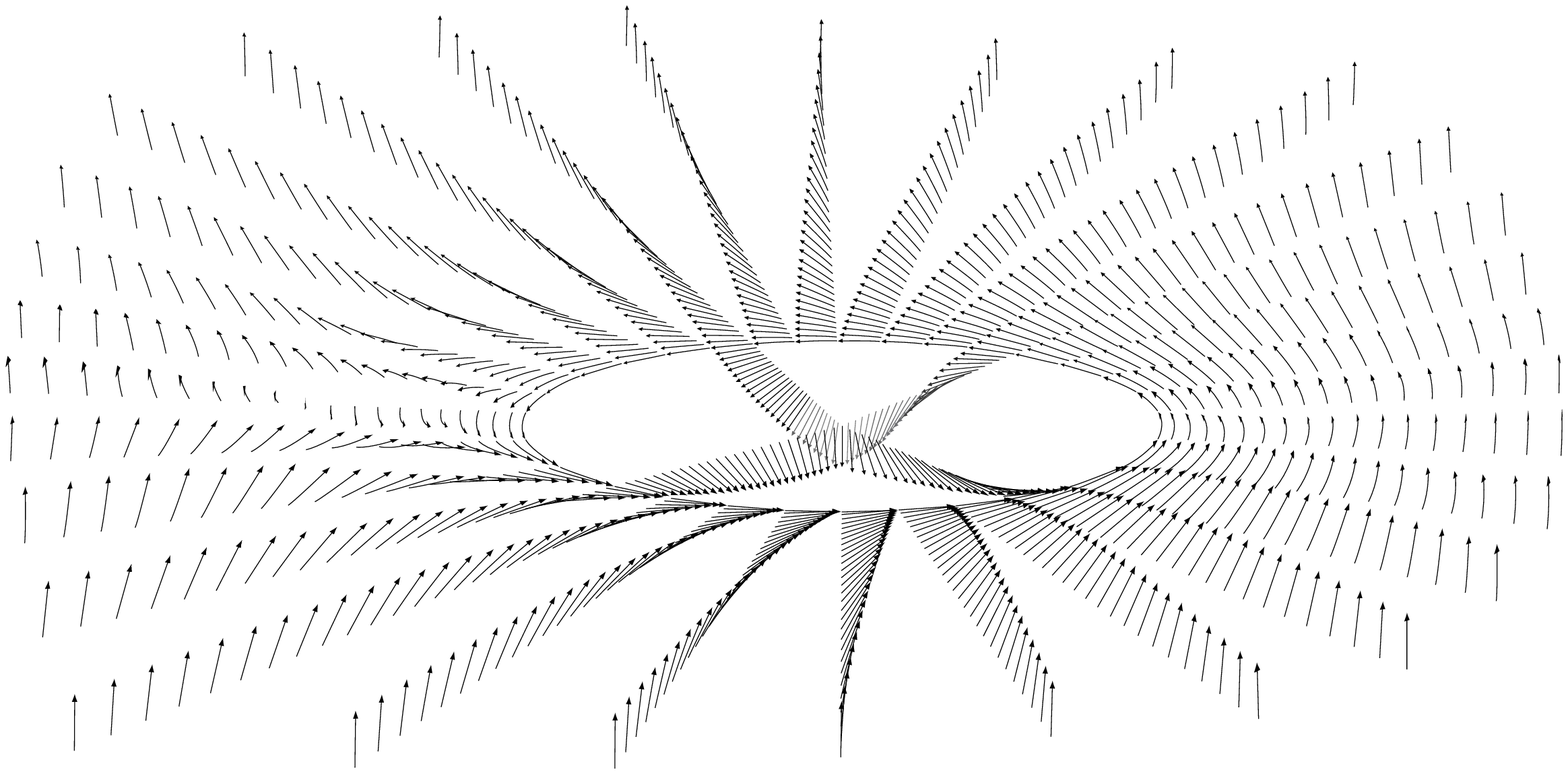}}
\caption{
A skyrmion spin texture in a quantum Hall ferromagnet.  The spins are
down at the origin but up at infinity.  At intermediate distances the
spins lie in the XY plane and have a vortex like configuration.  Even
though this object is an excitation of the bosonic spin field, the
quantized Hall coefficient causes it to
actually carry fermion number (and charge) equal to the quantum number
$\nu \equiv \frac{h}{e^2}\sigma_{xy}$.
}
\label{fig:skyrmion}
\end{figure}

With this energy functional, the system has neutral bosonic spin waves with
quadratic dispersion.  These spin waves represent small oscillations of the
spin orientation about the average direction.  Somewhat surprisingly, there
exist \cite{lee-kane,sondhi}
topological defects in the underlying vector boson field 
$\vec m(\vec r)$ which are
actually fermions (for the case of filling factor $\nu=1$).  (There is
an analogy here with the fermionic domain walls in the $S=1/2$ spin
chains.)
These charged objects are called skyrmions by analogy with
the topological defects which represent the nucleons in the Skyrme model of
nuclear physics.  Fig. (\ref{fig:skyrmion}) shows an example of such 
a spin texture.  The
spins are all pointing up at infinity and rotate smoothly downward towards 
the
origin.  At intermediate distances the spins lie in the XY  plane and have a
vortex-like arrangement.  Such strange objects exist in ordinary
two-dimensional ferromagnets (such as an iron film) but they are 
energetically
expensive and freeze out at low temperatures.  What is unique about the
quantum Hall ferromagnet is that it is an itinerant magnet with a precisely
quantized value of the Hall coefficient.  It turns out that this causes the
skyrmions to carry a precisely quantized charge and fermion number.  Thus by
adding or subtracting charge from the filled Landau level, one can force
skyrmions to exist even in the ground state of the system.

The size of a skyrmion is controlled by a competition between the Zeeman 
energy
which wants to shrink the number of over turned spins, and the Coulomb 
Hartree
energy which wants to spread the extra charge out over a large area.  At
ambient pressure in GaAs, a skyrmion contains about four flipped
spins \cite{barrett}.  Under high pressure the effective g  factor is driven
towards zero and the skyrmion spin can be as large as 
30 \cite{pressure-tuned}.
  The large ratio of spin to charge of skyrmions has been directly measured
using NMR \cite{barrett},  tilted field transport
measurements \cite{eisenstein-book} and  optically \cite{goldberg}.

In an ordinary ferromagnet the spin waves have a Zeeman gap which is much
larger than the characteristic nuclear Zeeman splitting.  Hence it is
impossible for the nuclei to emit or absorb electronic spin waves.  This
situation is altered dramatically in the presence of skyrmions.
The ground state of a quantum Hall ferromagnet slightly away from filling
factor $\nu=1$ is a lattice of skyrmions \cite{luis-lattice,prl-lattice}.
  This means that the system has
non-colinear order.  Associated with this is a new $U(1)$ degree of freedom in
which the spins rotate about the field direction.  In a colinear  magnet this
does not produce an excitation because all the spins are aligned with the
field.  Because skyrmions contain spins lying in the XY plane, such rotations
do represent new states.  Since the rotation is about the field direction,
this new collective mode does not have a Zeeman gap, but rather is a gapless
Goldstone mode much like that in a superfluid.  The analog of the phase of the
superfluid order parameter is the local azimuthal orientation angle of the
skyrmions.  The analog of the conjugate
boson number is the local number of flipped
spins.

Because this new collective mode is gapless it couples very strongly to the
nuclei and increases the relaxation rate $1/T_1$ by a factor of $10^3$ over the
zero magnetic field value.  This dramatic enhancement brings the nuclei into
thermal equilibrium which increases the specific heat \cite{bayot}
by factors exceeding $10^5$.

\section{Double-Layer Quantum Hall Systems}

Recent technological progress has allowed the construction of double-layer
quantum Hall systems \cite{eisenstein-book} and related wide single well
systems \cite{shayegan}
 in which two high-mobility electron gases are separated by only 
about $200\AA$, a distance comparable to the spacing between electrons.
Inter-layer correlations can therefore be comparable to intra-layer
correlations.  We have already seen that Coulomb exchange effects can produce
ferromagnetism.  In double layer systems there is a remarkable analog of this
effect involving inter-layer correlations.  Even if there is no tunneling
allowed between the layers, quantum mechanics permits the possibility that we
are uncertain which layer an electron is in.  To visualize the significance of
this, it is useful to define an `isospin' which is up when the electron is in
the left layer and down when the electron is in the right layer.  A coherent
superposition of these two possibilities is allowed just as is given in
eq.(\ref{eq:coherent}).
This of course represents an isospin state oriented in a direction given by the
angles $\theta$ and $\varphi$.  Just as for ordinary spin, Coulomb exchange
effects prefer for the isospin vectors to be locally parallel.  Because the
inter-layer interaction is slightly weaker than the intra-layer interaction,
the energy is not fully isospin rotation invariant but rather has the form
\begin{equation}
U = \frac{1}{2}\bar\rho_s\int d^2r\, \partial_\mu
m^\nu \partial_\mu m^\nu
 +\int d^2r\, \lambda (m^z)^2.
\end{equation}
The second term is an easy-plane anisotropy associated with the capacitive
charging energy.  If $m^z=\pm 1$, all of the electrons are in only one of the
layers and there is a large Coulomb cost.  This easy-plane anisotropy makes the
system into an XY model which has a Kosterlitz-Thouless phase transition into
an ordered state at low temperatures.  The long range order of this state has
been observed experimentally through the extreme sensitivity of the transport
to small tilts of the applied magnetic
field \cite{eisenstein-book,shayegan,smgbook}.
 The Kosterlitz-Thouless transition (which has not yet been directly 
observed)
is controlled by the unbinding of vortex-antivortex pairs, just as in a
superconducting or superfluid films.  As usual, an isolated vortex has a 
logarithmically divergent energy.  In a superconducting film
this is due to the kinetic energy stored
in the supercurrent circulating around the vortex.  In the present case it is
the loss of Coulomb exchange energy associated with the isospin currents
circulating around the vortex.

Recently a novel new type of broken symmetry involving a combination of isospin
and real spin has been proposed \cite{dassarma}
 to explain certain features of the Raman
spectrum in these systems \cite{pinczuk}.

\section{Summary}

In summary, we have seen that low-dimensional quantum systems can exhibit
novel types of order.  In some cases this order is hidden so that the 
system appears to be a featureless quantum liquid.  However there often 
exists a dual picture in which the system order appears naturally.  This 
dual picture can be achieved by interchanging the roles of charge and flux, 
particles and vortices, or waves and particles.  We have also seen that
the exotic ordering in 
low-dimensional quantum systems often produces new collective excitation 
modes and fractional quantum numbers.

{\ack This work was supported by NSF DMR 97-14055.
It is a pleasure to dedicate this paper to my long time friend and mentor,
Eli Burstein, on the occasion of his 80th birthday.}

\newpage

\newpage

\newpage

\newpage

\newpage

\newpage

\begin{thebibliography}{99}
\bibitem{spinon-expt} Tennant \textit{et al.}, \textit{Phys. Rev. Lett.}
\textbf{70}, 4003 (1993).

\bibitem{caveat}  Strictly speaking this order is perfect only in the 
so-called AKLT model which is closely related to the Heisenberg model.  
In the Heisenberg model, quantum fluctuations introduce pairs of defects 
in this order, but they are confined together and can never become widely 
separated.  Hence the two models are in the same phase and describe 
essentially the same physics.

\bibitem{dennijs} Rommelse, K. and den Nijs, M., \textit{Phys. Rev. Lett.}
\textbf{59}, 2578 (1987); den Nijs, M. and Rommelse, K., \textit{Phys. Rev.} B
\textbf{40}, 4709 (1989).

\bibitem{arovas-smg} Girvin, S.M. and Arovas, D.P., \textit{Physica Scripta}
\textbf{T27}, 156 (1989); Arovas, D.P. and Girvin, S.M., in \textit{Recent
Progress in Many-Body Theories} (Edited by T. L. Ainsworth, C. E. Campbell, B.
E. Clements, E. Krotscheck). Plenum, New York, 1992.

\bibitem{kohmoto} Kohmoto, M. and Tasaki, H., \textit{Phys. Rev.} B \textbf{46},
3486 (1992).

\bibitem{hymanthesis} Hyman, Ross A., Ph.D. Thesis (Indiana University, 1996),
unpublished.

\bibitem{hyman-yang} Hyman, R.A., Yang, Kun, Bhatt, R.N. and Girvin, S.M.,
\textit{Phys. Rev. Lett.} \textbf{76}, 839 (1996); Hyman, R.A. and Yang, Kun,
\textit{Phys. Rev. Lett.} \textbf{78}, 1783 (1997).

\bibitem{dassarmabook} \textit{Perspectives in Quantum Hall Effects}, (Edited by
Sankar Das Sarma and Aron Pinczuk). Wiley, New York, 1997.

\bibitem{smgbook} \textit{The Quantum Hall Effect}, 2nd Ed., (Edited by R. E.
Prange and S. M. Girvin). Springer, New York, 1990.

\bibitem{chakraborty} Chakraborty, T. and Pietil\"{a}inen, P., \textit{The
Fractional Quantum Hall Effect: Properties of an incompressible quantum fluid},
Springer Series in Solid State Sciences \textbf{85}, (Springer-Verlag Berlin,
Heidelberg, New York, Tokyo, 1988), and references therein.

\bibitem{stone} \textit{The Quantum Hall Effect}, (Edited by Michael Stone).
World Scientific, Singapore, 1992.

\bibitem{chap10} Girvin, S.M., Chap. 10 in Ref.~(\cite{smgbook}).

\bibitem{ODLRO} Girvin, S.M. and MacDonald, A.H., \textit{Phys. Rev. Lett.}
\textbf{58}, 1252 (1987).

\bibitem{Read} Read, N., \textit{Phys. Rev. Lett.} \textbf{62}, 86 (1988); Jain,
J.K. and Read, N., \textit{Phys. Rev.} B \textbf{40}, 2723 (1989).

\bibitem{Kivelson-Zhang} Zhang, S.C., Hansson, H. and Kivelson, S.,
\textit{Phys. Rev. Lett.} \textbf{62}, 82 (1989); \textbf{62}, 980 (1989).

\bibitem{wen} Wen, X.G., \textit{Int. J. Mod. Phys.} B \textbf{6}, 1711 (1992).

\bibitem{kane-fisher-book} Kane, C.L. and Fisher, M.P.A., in
Ref.~(\cite{dassarmabook}).


\bibitem{HLR} Halperin, B.I., Read, N., and Lee, P.A., \textit{Phys. Rev.} B
\textbf{47}, 7312 (1993).

\bibitem{jain-book-review} Jain, J., in Ref.~(\cite{dassarmabook}).

\bibitem{halperin-book-review} Halperin, B.I., in Ref.~(\cite{dassarmabook}).

\bibitem{fradkin} Lopez, A. and Fradkin, E., \textit{Phys. Rev.} B \textbf{44},
5246 (1991); \textbf{47}, 7080 (1993).

\bibitem{willett}  Willett, R. L., et al., \textit{Phys. Rev. Lett.}
\textbf{54}, 112 (1990); \textit{ Phys. Rev.} B \textbf{47}, 7344 (1993);
 \textit{Phys. Rev. Lett.} \textbf{71}, 3846 (1993).

\bibitem{kang} Kang, W., et al., \textit{Phys. Rev. Lett.} \textbf{71},
3850 (1993).

\bibitem{smg-ahmbook} For an elementary discussion of the physics of quantum
Hall ferromagnets see S.M. Girvin and A.H. MacDonald in
Ref.~(\cite{dassarmabook}).

\bibitem{lee-kane} Lee, D.H. and Kane, C.L., \textit{Phys. Rev. Lett.}
\textbf{64}, 1313 (1990).

\bibitem{sondhi} Sondhi, S.L., Karlhede, A., Kivelson, S.A. and Rezayi, E.H.,
\textit{Phys. Rev.} B \textbf{47} 16419 (1993).

\bibitem{barrett} Tycko, R., Barrett, S.E., Dabbagh, G., Pfeiffer, L.N. and
West, K.W., \textit{Science} \textbf{268}, 1460 (1995); Barrett, S.E., Dabbagh,
G., Pfeiffer, L.N., West, K.W. and Tycko, R., \textit{Phys. Rev. Lett.}
\textbf{74}, 5112 (1995).

\bibitem{pressure-tuned} Maude, D.K. \textit{et al.}, \textit{Phys. Rev. Lett.}
\textbf{77}, 4604 (1996); Leadley, D.R. \textit{et al.}, (LANL cond-mat/9706157,
unpublished).

\bibitem{eisenstein-book} Eisenstein, J.P. in Ref.~(\cite{dassarmabook}).

\bibitem{goldberg} Aifer, E.H., Goldberg, B.B. and Broido, D.A., \textit{Phys.
Rev. Lett.} \textbf{76}, 680 (1996).

\bibitem{luis-lattice} Brey, L., Fertig, H.A., C\^{o}t\'{e}, R. and MacDonald,
A.H., \textit{Phys. Rev. Lett.}, \textbf{75}, 2562 (1995); Fertig, H.A., Brey,
L., C\^{o}t\'{e}, R., MacDonald, A.H., Karlhede, A. and Sondhi, S.,
\textit{Phys. Rev.} B \textbf{55}, 10671 (1997).


\bibitem{prl-lattice} C\^{o}t\'{e}, R., MacDonald, A.H., Brey, L., Fertig, H.A.,
Girvin, S.M. and Stoof, H.T.C., \textit{Phys. Rev. Lett.} \textbf{78}, {4825}
(1997).

\bibitem{bayot} Bayot, V., Grivei, E., Melinte, S., Santos, M.B. and Shayegan,
M., \textit{Phys. Rev. Lett.} \textbf{76}, 4584 (1996); \textit{ibid.}
\textbf{79}, 1718 (1997).

\bibitem{shayegan} Santos, M.B., Engel, L.W., Hwang, S.W. and Shayegan, M.,
\textit{Phys. Rev.} B \textbf{44}, 5947 (1991); Lay, T.S., Suen, Y.W.,
Manoharan, H.C., Ying, X., Santos, M.B. and Shayegan, M., \textit{Phys. Rev.} B
\textbf{50}, 17725 (1994).

\bibitem{dassarma} Das Sarma, S., Sachdev, Subir and Zheng, Lian, \textit{Phys.
Rev. Lett.} \textbf{79}, 917 (1997).

\bibitem{pinczuk} Pellegrini, Vittorio, Pinczuk, Aron, Dennis, Brian S., Plaut,
Annette S., Pfeiffer, Loren N. and West, Ken W., \textit{Phys. Rev. Lett.}
\textbf{78}, 310 (1997). 
\end{thebibliography}
\end{document}